\documentclass[%
 reprint,
superscriptaddress,
nofootinbib,
 amsmath,amssymb,
 aps,
]{revtex4-1}

\usepackage{graphicx}
\usepackage{dcolumn}
\usepackage{bm}
\usepackage{multirow}
\usepackage{color}

\begin{document}

\preprint{APS/123-QED}

\title{Improved search for two body muon decay ${\mu}^+{\rightarrow}e^{+}X_H$}

\author{A.~Aguilar-Arevalo}
\affiliation{Instituto de Ciencias Nucleares, Universidad Nacional Aut\'onoma de M\'exico, CDMX 04510, M\'exico}

\author{ M.~Aoki}
\affiliation{Physics Department, Osaka University, Toyonaka, Osaka, 560-0043, Japan}

\author{M.~Blecher}
\affiliation{Virginia Tech., Blacksburg, VA, 24061, USA}

\author{D.I.~Britton}
\affiliation{SUPA - School of Physics and Astronomy, University of Glasgow, Glasgow, United Kingdom}

\author{D.~vom~Bruch}
\thanks{Present address: LPNHE, Sorbonne Universit\'e, Universit\'e Paris Diderot, CNRS/IN2P3, Paris, France.}
\affiliation{Department of Physics and Astronomy, University of British Columbia, Vancouver, B.C., V6T 1Z1, Canada}

\author{D.A.~Bryman}
\thanks{Corresponding author (doug@triumf.ca).}
\affiliation{Department of Physics and Astronomy, University of British Columbia, Vancouver, B.C., V6T 1Z1, Canada}
\affiliation{TRIUMF, 4004 Wesbrook Mall, Vancouver, B.C., V6T 2A3, Canada}

\author{S.~Chen}
\affiliation{Department of Engineering Physics, Tsinghua University, Beijing, 100084, China}

\author{J.~Comfort}
\affiliation{Physics Department, Arizona State University, Tempe, AZ 85287, USA}

\author{S.~Cuen-Rochin}
\affiliation{TRIUMF, 4004 Wesbrook Mall, Vancouver, B.C., V6T 2A3, Canada}
\affiliation{Universidad Aut\'onoma de Sinaloa, Culiac\'an, M\'exico}

\author{L.~Doria}
\affiliation{TRIUMF, 4004 Wesbrook Mall, Vancouver, B.C., V6T 2A3, Canada}
\affiliation{PRISMA$^+$ Cluster of Excellence and Institut f\"ur Kernphysik, Johannes Gutenberg-Universit\"at Mainz, Johann-Joachim-Becher-Weg 45, D 55128 Mainz, Germany}

\author{P.~Gumplinger}
\affiliation{TRIUMF, 4004 Wesbrook Mall, Vancouver, B.C., V6T 2A3, Canada}

\author{A.~Hussein}
\affiliation{TRIUMF, 4004 Wesbrook Mall, Vancouver, B.C., V6T 2A3, Canada}
\affiliation{University of Northern British Columbia, Prince George, B.C., V2N 4Z9, Canada}

\author{Y.~Igarashi}
\affiliation{KEK, 1-1 Oho, Tsukuba-shi, Ibaraki, Japan}

\author{S.~Ito}
\thanks{Corresponding author (s-ito@okayama-u.ac.jp).\\Present address: Faculty of Science, Okayama University, Okayama, 700-8530, Japan.}
\affiliation{Physics Department, Osaka University, Toyonaka, Osaka, 560-0043, Japan}

\author{S.~Kettell}
\affiliation{Brookhaven National Laboratory, Upton, NY, 11973-5000, USA}

\author{L.~Kurchaninov}
\affiliation{TRIUMF, 4004 Wesbrook Mall, Vancouver, B.C., V6T 2A3, Canada}

\author{L.S.~Littenberg}
\affiliation{Brookhaven National Laboratory, Upton, NY, 11973-5000, USA}

\author{C.~Malbrunot}
\thanks{Present address: Experimental Physics Department, CERN, Gen\`eve 23, CH-1211, Switzerland.}
\affiliation{Department of Physics and Astronomy, University of British Columbia, Vancouver, B.C., V6T 1Z1, Canada}

\author{R.E.~Mischke}
\affiliation{TRIUMF, 4004 Wesbrook Mall, Vancouver, B.C., V6T 2A3, Canada}

\author{T.~Numao}
\affiliation{TRIUMF, 4004 Wesbrook Mall, Vancouver, B.C., V6T 2A3, Canada}

\author{D.~Protopopescu}
\affiliation{SUPA - School of Physics and Astronomy, University of Glasgow, Glasgow, United Kingdom}

\author{A.~Sher}
\affiliation{TRIUMF, 4004 Wesbrook Mall, Vancouver, B.C., V6T 2A3, Canada}

\author{T.~Sullivan}
\thanks{Present address: Department of Physics, University of Victoria, Victoria BC V8P 5C2, Canada.}
\affiliation{Department of Physics and Astronomy, University of British Columbia, Vancouver, B.C., V6T 1Z1, Canada}

\author{D.~Vavilov}
\affiliation{Department of Physics and Astronomy, University of British Columbia, Vancouver, B.C., V6T 1Z1, Canada}
\affiliation{TRIUMF, 4004 Wesbrook Mall, Vancouver, B.C., V6T 2A3, Canada}


\collaboration{The PIENU Collaboration}



\date{\today}

\begin{abstract}

Charged lepton flavor violating muon decay ${\mu}^+{\to}e^+X_H$, where $X_H$ is a massive neutral boson, was sought by searching for extra peaks in the muon decay ${\mu}^+{\to}e^+{\nu}\bar{\nu}$ energy spectrum in the $m_{X_H}$ mass region $47.8-95.1$ MeV/$c^2$. 
No signal was found and 90\% confidence level upper limits were set on the branching ratio ${\Gamma}({\mu}^+{\to}e^+X_H)/{\Gamma}({\mu}^+{\to}e^+{\nu}\bar{\nu})$ at the level of $10^{-5}$ for this region. 

\end{abstract}

\maketitle


\section{\label{sec:Introduction}Introduction}

Observations of neutrino oscillations have established that lepton flavor is not strictly conserved. 
In the context of the Standard Model (SM), however, charged lepton flavor violating (CLFV) effects are too small to be observed \cite{CLFV}. 
Massive or massless weakly interacting neutral bosons $X$ such as axions  \cite{Axion1,Axion2,Axion3,Axion4} and majorons \cite{Majoron1,Majoron2,Majoron3} have been suggested to extend the SM including models with dark matter
candidates, baryogenesis, and solutions to the strong CP problem.
Wilczek suggested such a model \cite{Familon} which may lead to CLFV where the boson X can be emitted in flavor changing interactions.  
Such new bosons have been sought by experiments using kaon \cite{KXH1,KXH2,KX01,KX02,KX03, Hou}, pion \cite{SINDRUM,Picciotto}, and muon decays \cite{Derenzo, Doug, PSI, TWIST, Jodidio}. 

When decay products from a massive boson $X_H$ are not detected due to, for example, a long lifetime, CLFV two body muon decay involving a massive boson ${\mu}^+{\to}e^+X_H$ can be sought by searching for extra peaks in the muon decay  ${\mu}^+{\to}e^+{\nu}\bar{\nu}$ positron energy spectrum. 
The mass of the boson $m_{X_{H}}$ can be reconstructed using the equation
\begin{equation}\label{eq:mass}
m_{X_H}=\sqrt{m_{\mu}^2+m_e^2-2m_{\mu}E_e},
\end{equation}
where $m_{\mu}$ and $m_e$ are the masses of the muon and the positron, respectively, and $E_e$ is the total energy of the decay positron. 

\begin{figure}[]
\includegraphics[width=8.5cm]{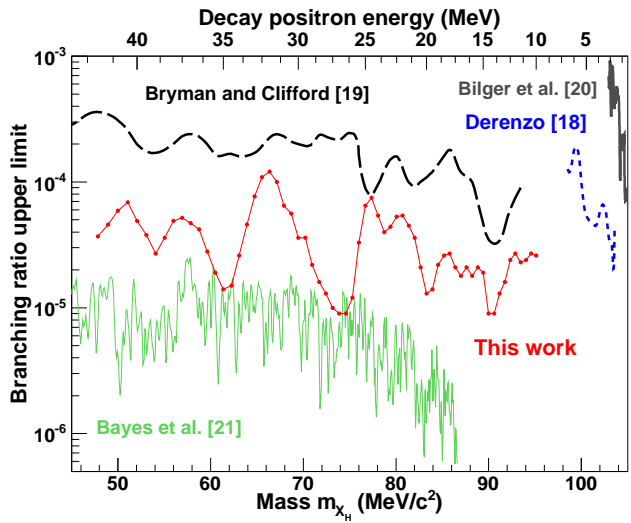}
\caption{\label{fig:FinalResult}Summary of the experimental upper limits on the ${\mu}^+{\rightarrow}e^+X_H$ branching ratio. The filled red circles with the thin solid red line show the results of this work. The limits represented by the dotted blue line, thick dashed black line, thick solid gray line, and thin solid green line are from Refs. \cite{TWIST, Doug, Derenzo, PSI}, respectively.}
\end{figure}

Two-body muon decays ${\mu}^+{\to}e^+X_H$ were searched for by Derenzo \cite{Derenzo} using a magnetic spectrometer; experimental limits\footnote{All limits quoted in this paper are at the 90\% confidence level.}  on the branching ratio ${\Gamma}({\mu}^+{\to}e^+X_H)/{\Gamma}({\mu}^+{\to}e^+{\nu}\bar{\nu})<2{\times}10^{-4}$  were set in the mass region from 98.1 to 103.5  MeV/$c^2$. 
Exotic muon decays were also sought as a byproduct of the ${\pi}^+{\rightarrow}e^+{\nu}$ branching ratio measurement \cite{Bryman} by Bryman and Clifford \cite{Doug} using a NaI(T$\ell$) calorimeter, resulting in upper limits on the branching ratio ${\lesssim}3{\times}10^{-4}$ in the mass range from 39.3 to 93.4  MeV/$c^2$. 
Muon decay in the mass region up to the kinetic limit was studied by Bilger {\it et al}. \cite{PSI} using a germanium detector. 
The most sensitive experiment done so far by Bayes {\it et al}. \cite{TWIST} gave limits from $10^{-5}$ to $10^{-6}$ in the mass range from 3.2 to 86.6 MeV/$c^2$. 
Figure \ref{fig:FinalResult} shows a summary of the present status of the search for  ${\mu}^+{\rightarrow}e^+X_H$ decay with upper limits in the mass region from 45 to 105 MeV/$c^2$. 
A massless boson $X_0$ was also searched for by Jodidio {\it et al}. \cite{Jodidio}, and the upper limit on the branching ratio was found to be ${\Gamma}({\mu}^+{\rightarrow}e^+X_0)/{\Gamma}({\mu}^+{\to}e^+{\nu}\bar{\nu})<2.6{\times}10^{-6}$.

The present work was carried out with data from the PIENU experiment principally designed to measure the  branching ratio ${\Gamma}[{\pi}^+{\to}e^+{\nu}({\gamma})]/{\Gamma}[{\pi}^+{\to}{\mu}^+{\nu}({\gamma})]$ using pion decays at rest \cite{PIENU}. 
A 75 MeV/$c$ ${\pi}^+$ beam from the TRIUMF M13 channel \cite{M13} was degraded by two thin plastic scintillator beam counters. Pion tracking was performed by two multiwire proportional chambers and two silicon strip detectors. 
The pion beam was stopped in an 8 mm thick plastic scintillator target. 
Positrons from ${\pi}^+{\to}{e}^+{\nu}$ decays and ${\mu}^+{\to}e^+{\nu}\bar{\nu}$ decays following ${\pi}^+{\rightarrow}{\mu}^+{\nu}$ decays were measured by two thin plastic scintillators used as telescope counters and a calorimeter consisting of a 48 cm (dia.) ${\times}$ 48 cm (length)  single crystal NaI(T$\ell$) detector surrounded by pure CsI crystals \cite{PIENUNIMA}. 
A silicon strip detector and a multiwire proportional chamber were used to reconstruct tracks of decay positrons and define the acceptance. 
The energy resolution of the calorimeter was 2.2\% (FWHM) for 70 MeV positrons. 
A total of $1.9{\times}10^8$ muon decays were used to search for the decay ${\mu}^+{\rightarrow}e^+X_H$ with lifetime ${\tau}_X>10^{-9}$ s. 
The energy resolution is a factor of two improvement and the statistics are an order of magnitude larger than the previous TRIUMF experiment \cite{Doug}. 
The present experiment is also sensitive to a higher mass region than that of Ref. \cite{TWIST}.

\section{Analysis}

\begin{figure}[]
\includegraphics[width=8.5cm]{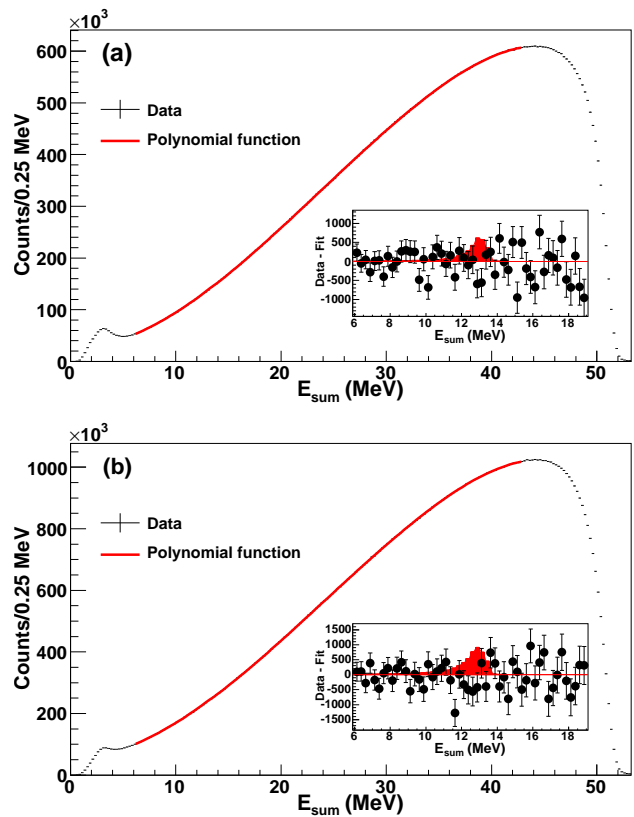}
\caption{\label{fig:MichelResi} The muon decay energy spectra from data taken before (a) and after (b) November, 2010 fit to polynomial functions (solid red line). The insert boxes show the residuals in the low energy region with statistical uncertainties (black circles) and a hypothetical signal (from MC) with the branching ratio $5.0{\times}10^{-5}$ with $m_{X_H}=90$ MeV/$c^2$ (red histograms). The bumps at 3 MeV were due to the low energy positrons that hit the telescope counters but did not reach the calorimeter (see text).}
\end{figure}

The data in the PIENU experiment were taken in runs occurring from 2009 to 2012. 
Because the energy calibration system for the CsI crystals was not available before November 2010, the data were divided into two sets, before and after that date. 
Pions were identified using energy loss information in the beam counters. 
Any events with extra hits in the beam and telescope counters were rejected. 
To ensure the events were from muon decay, the late time region $>200$ ns after the pion stop was selected. 
A solid angle cut of about 15\% was used for the data set after November 2010. 
A tighter acceptance cut (corresponding to about 10\% solid angle) was applied to the data taken before November 2010 to minimize electromagnetic shower leakage. 
Figure \ref{fig:MichelResi} shows the muon decay energy spectra for those two data sets where  $E_{\rm sum}$ is the sum of energies observed in the calorimeter, telescope counters, and silicon strip  detector including positron annihilation but excluding approximately 1.5 MeV energy loss in the target and inactive materials. 
The bumps at about 3 MeV in the low energy region of the spectra were due to positrons which hit the telescope counters but did not enter the calorimeter; positron annihilation in the last telescope scintillator resulted in one 0.511 MeV photon depositing energy in the calorimeter.

The two muon decay energy spectra were each fit to smooth 6th order polynomial functions in the energy region $E_{\rm sum}=6$ to $43$ MeV but excluding a region from -1.75 to +1.25 MeV around a possible signal peak where the search was to be performed. 
Then, for each $m_{X_H}$, the spectra were fit simultaneously to the polynomial functions with fixed fitting parameters obtained in the initial procedure plus a peak signal shape for the decay ${\mu}^+{\to}e^+X_H$. 
To combine the two data sets, a common branching ratio was used as a free parameter in the fit. 
The validity of the fit procedure was confirmed using the simulated muon decay energy spectrum and the signal peak with the branching ratio $1.0{\times}10^{-4}$ at several energies.
The polynomial function fit without any added signal shape resulted in ${\chi}^2/{\rm d.o.f}=1.09$ (${\rm d.o.f}=282$). 
The signal shapes were produced by a Monte Carlo (MC) simulation \cite{Geant4} that reproduced the peak of the decay ${\pi}^+{\rightarrow}e^+{\nu}$ at 69.8 MeV. 
This procedure was repeated in the range $E_{\rm sum}=$ 8.5 to 40.5 MeV (corresponding to the actual decay positron energy $E_{e}=$ 10 to 42 MeV) with 0.5 MeV steps.

\section{Results and Conclusion}

No extra peaks due to CLFV muon decay ${\mu}^+{\to}e^+X_H$ with a lifetime ${\tau}_{X}>10^{-9}$ s were observed and upper limits on the branching ratio ${\Gamma}({\mu}^+{\to}e^+X_H)/{\Gamma}({\mu}^+{\to}e^+{\nu}\bar{\nu})$ from $10^{-5}$ to $10^{-4}$ were set for the mass region $m_{X_H}=$ 47.8 to  95.1 MeV/$c^2$ as shown  in Fig. \ref{fig:FinalResult}. 
Statistics were the dominant source of uncertainty on the branching ratios. 
Systematic uncertainties and acceptance effects were approximately canceled by taking the ratio of the fit amplitude of signal events to the number of total muon decays. 
Improved and new limits in the mass region from 87.0 MeV/$c^2$ to 95.1 MeV/$c^2$ were set.

\begin{acknowledgments}
This work was supported by the Natural Sciences and Engineering Research Council of Canada (NSERC, number SAPPJ-2017-00033), and by the Research Fund for the Doctoral Program of Higher Education of China, by CONACYT doctoral fellowship from Mexico, and by JSPS KAKENHI Grant numbers 18540274, 21340059, 24224006, 17H01128, 19K03888 in Japan. 
We are grateful to Brookhaven National Laboratory for the loan of the crystals, and to the TRIUMF operations, detector, electronics and DAQ groups for their engineering and technical support.

We would also like to thank to R. Bayes and A. Olin for providing the experimental data in Ref. \cite{TWIST}. 
\end{acknowledgments}

\end{document}